\documentclass
{mn2e}
\usepackage{epsfig}
\usepackage{amsmath, amssymb,bm}
\usepackage{xcolor}

\def \be{\begin{equation}}
\def \ee{\end{equation}}

\def \msun{\rm M_{\odot}}

\title[Angular Momentum Transfer in QPEs]{Angular Momentum Transfer in QPEs from Galaxy Nuclei}

\author[Andrew King]{
Andrew King$^{1,2,3, \star}$ \\
  $^{1}$ Department of Physics \& Astronomy, University of Leicester, Leicester, LE1 7RH, UK \\
  $^{2}$ Astronomical Institute Anton Pannekoek, University of Amsterdam, Science Park 904, 1098 XH Amsterdam, Netherlands\\ 
  $^{3}$ Leiden Observatory, Leiden University, Niels Bohrweg 2, NL-2333 CA Leiden, Netherlands \\
  $^{\star}$ {E-mail:~} {\rm ark@astro.le.ac.uk} }

\date{Accepted XXX. Received YYY; in original form ZZZ}

\pubyear{2019}

\begin{document}
\label{firstpage}
\pagerange{\pageref{firstpage}--\pageref{lastpage}}
\maketitle

\begin{abstract}
A suggested model for quasi--periodic eruptions (QPEs) from galaxy nuclei invokes a white dwarf in an eccentric orbit about the central massive black hole. I point out that the extreme mass ratio allows the presence of strong Lindblad resonances in the accretion disc. These are important for the stability of mass transfer, and may trigger the eruptions themselves by rapidly transferring angular momentum from the accretion disc (which is likely to be eccentric itself) to the orbiting WD companion at pericentre. I consider the effects of
von Zeipel--Lidov--Kozai (ZLK) cycles caused by a perturber on a more distant orbit about the central black hole. 
I show that ZLK cycles can change the orbital periods of
QPE systems on timescales much shorter than the mass transfer time, as seen in ASASSN-14ko, and produce
correlated short--term variations of their mass transfer rates and orbital periods, as recently observed in GSN~069. 
Further monitoring of these sources 
should constrain the parameters of 
any perturbing companions. This in turn may constrain the nature of the events creating QPE systems, and perhaps give major insights into how the central black holes in low--mass galaxies are able to grow.

\end{abstract}




\section{Introduction} \label{sec:intro}

X--ray observations of the nuclei of 
low--mass galaxies show that several of them
produce quasi--periodic eruptions (QPEs: Miniutti et al., 2019; Giustini et al. 2020; Song et al. 2020;
Arcodia et al. 2021; Chakraborty et al. 2021; Payne et al., 2021, 2022). 
Typically these sources have outbursts by factors $\sim 100$ in X--rays, 
%
which recur in roughly periodic fashion. The recurrence times currently range from a few hours
up to $\sim 100$~d or $\sim  1$~yr, and it is likely that these limits will widen as data accumulate from direct observations and archival searches. 
The X--rays have ultrasoft blackbody spectra 
and luminosities 
implying typical radii
$\gtrsim R_g = GM_1/c^2 = 6 \times 10^{10}m_5$~cm, of order the gravitational ($R_g$) and ISCO radii of a black hole of mass $M_1 \sim 10^5m_5\msun$.  These are consistent with the massive black holes (MBHs) one might expect in these low--mass galaxy nuclei.

In (King, 2020) I suggested a simple model for the first known QPE source GSN 069. This postulated
a low--mass star (with mass $M_2 \ll M_1$) -- found from the requirement of internal consistency with observational selection to be a white dwarf (WD) -- on a highly eccentric orbit about the central MBH, transferring mass to it at pericentre passage via an accretion disc. In a second paper (King, 2022, hereafter K22) I used the parametrization introduced by Chen et al. (2022), extended to the full WD mass range, to show that this model
was consistent with the data for 5 of the 6 known QPE sources, together with a previously unassociated system HLX-1, where the periodicity is $\sim 1$~yr. (I discuss the `missing' system ASSASN--14ko in Note 1 to Table 1 at the end of the paper.)

In all cases, loss of orbital energy $E$ and angular momentum $J$ via gravitational radiation (GR) is the ultimate driver of mass transfer, and observational selection effects mean that the orbiting star is in practice a low--mass white dwarf (WD) in detectable QPE systems (K22). 
This is the likely explanation for the presence of CNO--processed material found in GSN~069 (Sheng et al., 2021).

This model requires that
mass transfer is dynamically stable; the Roche lobe and the WD radius must move in step as GR reduces the orbital semi--major axis $a$ and the eccentricity $e$ on the same timescale. Since the WD expands as it loses mass, it must gain angular momentum from the accretion disc
and move in a wider orbit, implying a larger tidal radius. 

This stability has recently been questioned, so I discuss it further in Section 2.  
Orbital resonances within the accretion disc are likely to cause the required enhanced angular momentum transfer to the WD. 
The resonances may also directly cause the quasiperiodic eruptions themselves as the orbiting WD passes pericentre -- 
a similar origin has been suggested for the superoutbursts of the stellar--mass SU UMa binaries (Osaki \& Kato, 2013).

During the orbital evolution of a QPE binary the pericentre separation 
$p = a(1-e)$ remains almost constant. This is reasonable, since significant GR effects only occur in an effective point interaction at pericentre. 
Unusually, the interaction in QPE systems is strong enough that the resulting instantaneous mass transfer rate is  close to the long--term evolutionary mean driven by GR.
This is very different from the situation in accreting stellar--mass binaries. There the instantaneous rate oscillates widely around the evolutionary mean because of unrelated short--term effects. It only converges to this mean when averaged on timescales long compared with that taken for the driving process (here GR) to move the critical (Roche) lobe by a distance of order the density scaleheight of the donor star. This means for example that it is generally unsafe to try to deduce the mass transfer rate of a stellar--mass binary by measuring the rate of change of its orbital period -- see King \& Lasota (2021) for a recent discussion -- whereas QPE sources appear to be constrained to stay close to this mean mass transfer rate.

Although the  simple model of King (2020) and K22 works well for QPE systems, there is clearly more complexity in the structure of these sources. The
QPE system ASASSN-14ko (Payne et al., 2021, 2022) 
has a very nearly strict period of $P = 114$~days, and its mass transfer rate and luminosity agree with the predictions of GR driving. But the predicted GR period derivative
  $\dot P \simeq  - 1.3 \times 10^{-6}$  is in flat contradiction with
the measured value $\dot P = - 1.7\times 10^{-3}$ (Payne et al., 2021).

A second deviation from the expectations of K22 emerges from the 
recent thorough observational study by Minutti et al. (2022) of the first QPE source, GSN~069, which gives a historical X--ray light curve. This shows an episode where the orbital period appears to increase by a factor $\sim 1.3$ over a timescale of order 3000~d, and the mass transfer rate declines significantly on the same timescale.

In this paper I suggest that the underlying cause of the unusual behaviour of both ASASSN-14ko and GSN~069 is that in each of these two galaxies the central MBH--WD binary system is not isolated, but gravitationally influenced by a perturber which itself is in a wider orbit about the MBH. This object may be a star (or star cluster) which was part of the infall event causing the formation of the inner QPE `binary' itself. The interaction between the inner and outer binaries induces a variety of effects, now collectively known as von Zeipel--Lidov--Kozai (ZLK) cycles\footnote{The author ordering ZLK I adopt here follows the historical sequence in which the authors studied the interaction between two binaries in the contexts of the Solar System (von Zeipel 1910; Kozai, 1962), and  
artificial satellites in the Earth--Moon system (Lidov, 1961, 1962).},
 usually studied in the case where the secondary component in the inner binary (here the WD) has negligible mass compared with the primary (MBH) and the perturber. In many cases one can treat the problem by expanding the combined gravitational potential only to quadrupole order, effectively modelling the two binaries as wire loops with the masses spread around their orbits. As I explicitly noted in K22, it is inherently plausible that QPE binaries should be accompanied by other more distant orbiting stars (or star clusters), as such satellites of the MBH are likely additional results of the tidal capture events which probably produce QPE binaries (cf Cufari, Coughlin \& Nixon, 2022).

I show here that ZLK cycles can account for large orbital period derivatives, as seen in ASASSN-14ko, which are unrelated to the mass transfer process, and can also cause correlated
rapid changes change of orbital period and mass transfer rate, as in GSN 069.  In general the parameter space open to a perturber is large in any given case, but the possibility of narrowing it down should encourage continued monitoring of QPE sources, as this may give insight into the capture events forming them.

\section{Mass Transfer Stability in QPE Binaries, and the Origin of the 
Eruptions}

The QPE model discussed here requires that 
that mass transfer is dynamically stable, i.e. that the tidal lobe and the donor radius move in step as mass is transferred.
Suggestions to the contrary have appeared in the literature, but several of these neglected the orbital expansion driven by the tidal coupling of the accretion disc angular momentum when mass is transferred from the less massive star (here the WD) to a more massive accretor (the MBH); see K22 for a discussion. More recently, Lu \& Quataert (2022) have argued that in a highly eccentric system such as the QPE binaries considered here, the tidal coupling would actually transfer angular momentum from the donor to the accretion disc. This would cause mass transfer from a donor which expands on mass loss (as here) to be dynamically unstable, and the binary would coalesce in a few orbits.

The  argument of Lu \& Quataert (2022)
implicitly assumes that the accretion disc is circular. In that case matter at the outer edge of the
disc moves more slowly than the donor in an eccentric binary, and indeed the angular momentum transfer is from the donor to the disc, leading to instability. But 
it is rather unlikely that the disc is at all circular in QPE systems. Since the mass ratio $M_2/M_1$ is  $\ll 1$, the disc can easily grow large enough to contain strong Lindblad resonances (see e.g. Fig. 1 of Whitehurst \& King, 1991). These make the disc very eccentric and cause prograde apsidal precession, and are the cause of the superhump modulations at periods slightly longer than orbital in superoutbursts of the 
short--period ($P \lesssim 100~{\rm min}$)
SU UMa cataclysmic binaries (Lubow, 1991).
Early Lagrangian (e.g. SPH) simulations had readily found these effects, beginning with  Whitehurst (1988). Eulerian simulations took longer to achieve this, as the use of axisymmetric coordinates tends to suppress the disc eccentricity which is the basis of superhumps, but with attention to this problem superhumps now appear in these simulations too (Wienkers \& Ogilvie, 2018, and references therein).

SU UMa superhumps are driven by the 3:1 commensurability. In QPE binaries the much smaller mass ratios mean that the stronger 2:1 resonance is also accessible, so we can expect their discs to be strongly eccentric and precessing. In the SU UMa systems, superhumps appear during superoutbursts, when the discs undergo outbursts which are longer and brighter than their usual dwarf nova outbursts. A suggestive possibility (Osaki \& Kato, 2013) is that the tidal effects themselves actually cause the increased disc accretion of superoutbursts. By analogy, the presence of resonances in eccentric QPE sources may trigger the eruptions themselves when the donor is near pericentre. In addition, the precession of the eccentric disc would naturally cause deviations from strict periodicity, particularly in short--period systems. (I note that in long--period QPE systems such as ASSASN-14ko and HLX--1 the eruptions tend to be more regular; K22 discusses why in HLX--1 the disc may occasionally not re--form in time for the next periastron passage, and so miss an entire cycle). In this picture the angular momentum lost by the rapidly--accreting disc material is transferred to the WD orbit, maintaining orbital stability.

Simulations (which are presumably easier with SPH) are needed to check these suggestions, and in particular to determine the size and direction of angular momentum exchange between the eccentric binary orbit and the eccentric precessing disc. The presence of very significant CNO enhancement in at least one QPE source (Sheng et al., 2021) strongly supports the suggestion of WD donors in QPE sources, and so the idea that mass transfer is stable in them.

\section{ZLK cycles}

As remarked above, there is good reason to suspect the action of ZLK effects in QPE sources.
The characteristic feature of ZLK cycles is that the inner binary (the QPE system in our case) continuously exchanges its eccentricity $e$ with the inclination 
$i$ of its orbit to that of the outer (perturbing) binary. (The plane of the latter is almost fixed in many cases of interest, as the outer binary has the largest component of the whole system's total angular momentum.) The exchanges are subject to the constraint
\be
(1 - e^2)^{1/2}\cos i \simeq C ,
\label{am}
\ee
where $C$ is a constant set by the initial conditions.
This asserts that ZLK cycles have no effect on the angular momentum component of the inner binary orthogonal to its instantaneous plane. This is precisely the angular momentum $J$ being depleted by GR to drive mass transfer.

For given initial conditions the inner binary plane either librates (oscillates between two fixed inclinations $i$) or circulates (revolves continuously wrt the outer binary). The characteristic timescale for these motions is 
\be
t_{\rm ZLK} \simeq  \frac{8}{15\pi}\left(1 + \frac{M_1}{M_3}\right)\left(\frac{P_{\rm out}^2}{P}\right)(1 - e_{\rm out}^2)^{3/2},
\label{zlk}
\ee 
(Antognini, 2015),
where $M_1, M_3$ are the MBH and perturber masses  $P, P_{\rm out}$ are the periods of the inner (QPE) binary and the outer one respectively, and $e_{\rm out}$ is the eccentricity of the outer binary. We see that this timescale tends to infinity in the limit of vanishing perturber mass $M_3$, as expected.

ZLK cycles are quickly washed out if the binary precesses too rapidly, as this gradually destroys the near-resonance allowing the exchange of inclination and eccentricity. The strongest precession in QPE systems with WD donors (cf Willems, Deloye \& Kalogera, 2010) is the 
general--relativistic advance of pericentre, with period
\be
P_{\rm GR} = 4.27M_5^{-2/3}P_4^{5/3}(1 - e^2)\,{\rm yr},
\ee
so that
\be
\frac{P_{\rm GR}}{P} = 1.28\times 10^4M_5^{-2/3}
               P_4^{2/3}(1 - e^2),
\ee
where $e$ is the eccentricity, $M_5 = M_1/10^5\msun$ with $M_1$ the black hole mass, and $P_4$ is the orbital period $P$ in units of $10^4$~s.
This ratio is given for the current known systems in Table 1, and is always 
significantly larger than unity, although only of order 22 and 36 for two systems. This suggests that ZLK cycles can appear stably in most QPE systems, but may (if they appear at all) be fairly shortlived in some cases. I discuss this point further below.

\section{Evolution of the Inner binary during ZLK cycles}

ZLK cycles modulate the eccentricity $e$ of the inner (QPE) binary. But we see from (\ref{am}) that they have essentially no direct effect on its orbital angular momentum $J$. Since mass transfer is stable in a QPE binary, this system must evidently respond to ZLK changes of $e$ by holding constant its tidal lobe $R_2$: this must remain equal to the current radius of the WD donor, whose mass is unchanged. The lobe radius $R_2$ is proportional to the pericentre separation (see K22)
\be
p = a(1 - e),
\label{p}
\ee
so that the ZKL effect makes the semi--major axis $a$ change as
\be 
a\propto (1-e)^{-1}.
\label{a}
\ee
This in turn implies that ZLK cycles cause the period of a stably mass--transferring binary system to evolve as
\be 
P \propto a^{3/2} \propto (1-e)^{-3/2}.
\label{P}
\ee
The GR--driven mass transfer rate must evolve in response to the changes in $e$ and $P$ as 
\be
 -\dot M_2 \propto P^{-8/3}(1 - e)^{-5/2}
\propto P^{-1},
\label{pmdot}
\ee
where I have used eqn (15) of K22 together with (\ref{a}, \ref{P}).

The constraint (\ref{am}) implies that during a ZKL cycle the eccentricity $e$ reaches a maximum as the inner binary plane crosses the plane of the perturbing outer binary at $i = 0$.
From (\ref{P}) the inner binary period reaches a maximum at this point. Logarithmically differentiating (\ref{am}) 
we get
\be
\frac{e\dot e}{1-e^2} \simeq -\tan i \frac{{\rm d} i}{{\rm d} t}.
\ee
From this equation and (\ref{P}) we have
\be
\frac{\dot P}{P} = \frac{3}{2}\frac{\dot e}{1 - e} = 
-3\frac{1+e}{2e}\tan i\frac{{\rm d} i}{{\rm d} t}.
\label{dotp}
\ee
In all cases we expect ${\rm d} i/{\rm d}t \sim \pm 1/t_{\rm ZLK}$.

\section{Comparison with Observations}

\subsection{Period Changes}
We have seen that ZLK cycles can produce very rapid period changes
in QPE binaries (cf eqn \ref{dotp}), which may be accompanied by 
significant changes in the accretion luminosity (cf eqn \ref{pmdot}).
Because ZLK cycles produce these changes by altering the eccentricity affecting the GR losses driving mass transfer, there is no paradox in changes more rapid than given by the timescale $t_{\rm GR}$ for the latter. The apparently discordantly large period derivative of ASASSN--14ko is then a potential signature of this effect.
There are several ways to explain values of order the $\dot P = -1.7\times 10^{-3}$ observed there as a 
result of ZLK cycles.


If $\tan i \sim 1$ (i.e. the QPE plane is not close to the perturber plane)
we must have $e$ significantly smaller than unity. We see from Table 1
that this explanation cannot work for ASSASN--14ko itself, or any of the known QPE systems, which all have a much higher eccentricities.

Future observations may reveal QPE systems with lower $e$, and these would have 
$-\dot P \sim 3P/2et_{\rm ZLK}$, so from (\ref{zlk}) we find a value $\dot P \sim 10^{-3}$ would result if
the perturber mass and period are connected by 
\be
M_3\simeq 13\frac{em_5}{1+e}\left(\frac{P_{\rm out}}{P}\right)^2(1 - e_{\rm out}^2)^{3/2}\msun,
\ee
where $m_5 = M_1/10^5\msun$. We need $P_{\rm out}>P = 114\ {\rm d}$ for consistency in ASSASN--14ko. This is evidently possible with perturber having a normal stellar mass, as $e$ is very close to unity (see Table 1).

So we must look to other candidates for the perturbers in known QPE systems. Other possibilities are that the perturber is a star cluster rather than a single star, or that it is a single star with  extreme $e_{\rm out}$ approaching unity\footnote{In this case we have the {\it eccentric} ZLK effect: this becomes considerably more complicated, as now there are octupole contributions to the gravitational potential of similar order to the quadrupole ones considered so far. See Naoz (2016) for a review.}. This latter case may be more likely if 
the QPE binary and its perturber result from the same tidal capture event. This is important in highlighting the potential the QPE sources have in signalling these events, and their possible role in promoting black hole growth.
Clearly, only further observational monitoring of the known QPE systems can distinguish between all these possibilities.


\subsection{Light Curve and  Correlated Period Changes}

Equation (\ref{pmdot}) shows that ZKL cycles can continuously modify the mass transfer rate and so the luminosity of a QPE binary,
as recently observed in GSN~069 by Miniutti et al. (2022). As the period is increasing here, and the luminosity decreasing, we must have increasing eccentricity, so the plane of the QPE binary is approaching the perturber plane. These events should eventually appear in 
time--reversed order. The 3000~d timescale is easy to accomodate (cf eqn \ref{zlk}) with a stellar--mass perturber and an outer period not much longer than that of the QPE binary.

Presumably systems showing little change between eruptions, and no very rapid period changes, must either have no associated perturber, or a perturber period which is very long. Orbital changes induced by ZKL cycles might trigger other light curve effects, e.g. by cyclically altering disc accretion. These would add to the effect already noted by 
K22 that for systems with periods $P \gtrsim 1$~yr the accretion disc may have to re--form after a few outbursts, which may account for the missing outburst in HLX-1.

\section{Conclusions}

I have argued that the presence of resonances in the accretion disc makes it likely that in systems where a white dwarf orbits a massive black hole, mass transfer driven by the loss of gravitational wave energy is stable on a dynamical timescale. The resonances may also promote the rapid loss of disc angular momentum to the WD, and so directly cause the quasiperiodic eruptions. 

I have considered some of the effects that may appear in QPE systems because of von Zeipel--Lidov--Kozai (ZLK) cycles triggered by a perturber on a more distant orbit about the central massive black hole. The presence of perturbers of this kind appears likely, as they may be products of the same tidal capture events that formed the QPE binaries themselves.  Evidently more observations are needed to check the validity of the ZLK idea. If it is tenable, the parameter space available to the perturbers is currently very large, and still more observations would be needed to narrow it down.

QPE systems showing orbital period changes on timescales much shorter than the mass transfer time are obvious candidates for ZLK effects, and are very likely to reward further monitoring or archival searches.  For example, the predicted timescale for the disappearance of the ZLK cycles in ASSASN--14ko is only of order a decade. Similarly, correlated short--term variations of mass transfer rates and orbital periods in QPE systems may result from ZLK cycles. Here we can expect the data and interpretation to be more complex than for period changes, because other effects can also modulate the mass transfer rates. But this kind of study can potentially give major insights into how the central black holes in low--mass galaxies are able to grow.

\section*{Data Availability}

No new data were generated or analysed in support of this research.

\section*{Acknowledgments}

I thank Giovanni Miniutti for giving me early insight into important observational data and for many helpful and continuing discussions, and Chris Nixon and the anonymous referee for very helpful comments.

\begin{table*}
\caption{Parameters of the Current QPE Sample}
\vskip 5.0pt
\centering
{
\setlength{\tabcolsep}{3pt}
{
\hfill{}
  \begin{tabular}{|l||c|c|c|c|c|c|c|} 
    \hline
    Source & $P_4$ & $m_5$ & $(L\Delta t)_{45}$ & $m_2$   & $1-e$ & $P_{\rm GR}/P$
      \\
    \hline \hline
    eRO -- QPE2 & 0.86&  2.5    & 0.8 & 0.18&$9.9\times 10^{-2}$ & 1250
    \\
    XMMSL1 & 0.90 &  0.85    & 0.34 & 0.18 & $9.9\times 10^{-2}$ &2503      \\ 
      RXJ1301.9 &1.65& 18   &   1.7     & 0.15 & $7.2\times 10^{-2}$ & 36 
    \\
    GSN 069 & 3.16&  4.0 &  10 & 0.32& $2.8\times 10^{-2}$ & 130
    \\ 
     eRO-- QPE1 & 6.66 &9.1& 0.045 & 0.46 & $1.4\times 10^{-2}$& 291
     \\
     ASSASN--14ko &937& 700 & $3388$ & 0.56 & 
     $9.0\times 10^{-3}$ & 22 
     \\
     HLX--1 & 2000 & [0.5]& 1000 & 1.43 & $1.2\times 10^{-4}$ & 774
     \\
   \hline\hline
  \end{tabular}}
  \hfill{}
  }
  \vskip 0.2truecm
  \begin{itemize}
\item[] {\bf Note 1} 
This table is adapted from Table 1 of K22, but now ordered by period $P$. We note the general tendency that the eccentricity $e$ is smaller for shorter periods, consistent with the effects of GR losses. The very bright QPE source ASSASN-14ko (Payne et al., 2021, 2022) was missing from Table 1 of K22. The difficulty in modelling it arose because it is (uniquely) extremely close to the limit
\be
m_5(L\Delta t)_{45}^{1/3} \lesssim 10^4
\label{max}
\ee
required to avoid the model formally predicting that the WD pericentre distance $a(1-e)$ is larger than the innermost stable orbital radius, which is itself $\simeq R_g = GM_1/c^2$, where $R_g$ is the black hole's gravitational radius.
Here $m_5 = M_1/10^5\msun$, and $(L\Delta t)_{45}$ is the total energy radiated at pericentre passage in units of $10^{45}$~erg.  Equation (\ref{max}) is the condition
\be
a(1 - e)^3 > \left(\frac{GM_1}{c^2}\right)^3
\ee
written using the parametrization of Chen et al. (2022) followed in K22. 
Evidently the form (\ref{max}) expresses the facts that the radiated energy is increased in a tighter orbit, but that a larger black hole mass increases $R_g$.
For ASSASN-14ko we have $m_5 \simeq 700$, requiring $(L\Delta t)_{45} \lesssim 3388$, as compared with rough observational estimates $(L\Delta t)_{45} \simeq 4000$. Here I adopt the extreme value
$(L\Delta t)_{45} \lesssim 3388$ for this system.
For all other currently known QPE systems the constraint (\ref{max}) is very easily satisfied.
\item[]{\bf Note 2}
There is no secure mass estimate for the black hole in HLX--1. Here I adopt the minimum value $m_5 = 0.5$ allowing the donor to be below the Chandrasekhar mass (i.e. $m_2 \simeq 1.4$; see K22). Larger $m_5$ values allow smaller $m_2$ (see K22). 
\item[]{\bf Note 3}
The Table also gives the values of 
$P_{\rm GR}/P$  specifying the stability of possible ZLK cycles.
\end{itemize}
\end{table*}



\label{lastpage}
\end{document}